# Electric-field–tuned consecutive topological phase transitions between distinct correlated insulators in moiré MoTe$_2$/WSe$_2$ heterobilayer

Xumin Chang[1*], Zui Tao[2*], Bowen Shen[2], Wanghao Tian[1], Jenny Hu[3], Kateryna Pistunova[3], Kenji Watanabe[4], Takashi Taniguchi[4], Tony F. Heinz[3,5], Tingxin Li[1,6], Kin Fai Mak[2,7,8,9], Jie Shan[2,7,8,9], and Shengwei Jiang[1,6†]

[1] State Key Laboratory of Micro-nano Engineering Science, Key Laboratory of Artificial Structures and Quantum Control (Ministry of Education), School of Physics and Astronomy, Shanghai Jiao Tong University, Shanghai 200240, China
[2] School of Applied and Engineering Physics, Cornell University, Ithaca, NY, USA
[3] Departments of Physics and Applied Physics, Stanford University, Stanford, California, USA
[4] National Institute for Materials Science, 1-1 Namiki, Tsukuba 305-0044, Japan
[5] SLAC National Accelerator Laboratory, Menlo Park, California, USA
[6] Tsung-Dao Lee Institute, Shanghai Jiao Tong University, Shanghai, 201210, China
[7] Laboratory of Atomic and Solid State Physics, Cornell University, Ithaca, NY, USA
[8] Kavli Institute at Cornell for Nanoscale Science, Ithaca, NY, USA
[9] Max Planck Institute for the Structure and Dynamics of Matter, Hamburg, Germany

**ABSTRACT**
Consecutive topological phase transitions (TPTs) between strongly correlated electronic phases that differ simultaneously in symmetry breaking and topological order are of fundamental interest in condensed matter physics, yet, rarely realized experimentally. We report two consecutive electric-field–driven TPTs at half filling ($v = 1$) in angle-aligned MoTe$_2$/WSe$_2$ moiré heterobilayers. With increasing out-of-plane displacement field, a geometrically frustrated Mott insulator evolves into a ferromagnetic quantum anomalous Hall (QAH) Mott insulator, i.e., a spin-polarized topological Mott insulator without an observable charge-gap closure, and subsequently into an antiferromagnetic, valley-coherent Mott insulator (VC-AFM) accompanied by a continuous charge-gap collapse and the emergence of a critical metallic state. Layer-resolved magnetic circular dichroism (MCD), magneto-transport, and compressibility measurements jointly determine the phase diagram. The high-field evolution of the antiferromagnetic state reveals a metamagnetic-like transition at a critical field $B^*$, above which a Chern insulating transport response reappears. Our results establish the MoTe$_2$/WSe$_2$ moiré platform as a tunable realization of an extended Kane–Mele–Hubbard model hosting sequential correlation–topology intertwined transitions.

## I. INTRODUCTION.

Condensed phases of matter are categorized by their symmetry and topology [1,2]. Understanding the quantum phase transition between states with distinct symmetries and topological orders is one of the forefronts in condensed matter physics [3–5]. Whereas TPTs between non-interacting states of matter (e.g., topological to band insulator transition) have been extensively studied [1,2], TPTs between strongly correlated states of matter remain largely unexplored [6,7], at least in the experimental realm. Transition metal dichalcogenide (TMD) semiconductor moiré materials have emerged as a promising platform to study exotic TPTs involving strong electronic correlations. In this class of materials, the extended Hubbard model captures the low-energy physics when the flat bands are isolated and non-topological [6,8–10]; a suite of correlated phenomena, including Mott insulator, generalized Wigner crystals, ferromagnetism, and superconductivity, has been demonstrated [11–20]. When two nearby flat bands are intertwined, topological moiré bands with non-zero valley-resolved Chern number emerge, realizing the Kane-Mele-Hubbard model with a staggered sublattice potential in a honeycomb lattice [21–23,7,24,25]. The combined effects of electronic correlations and band topology in this model have been demonstrated to stabilize various spin-valley-symmetry breaking ground states with different topological orders, including integer and fractional QAH [7,18,21,23,26–33]. The system provides a unique platform to study quantum phase transitions between these states in an intrinsic manner by tuning band topology and correlation.

In this study, we perform comprehensive magneto-transport and optical studies on the TPTs between distinct Mott insulators in angle-aligned MoTe$_2$/WSe$_2$ moiré bilayers. We examine dual-gated Hall bar devices. The 7 % lattice mismatch between MoTe$_2$ and WSe$_2$ gives a high moiré density $\sim 5 \times 10^{12}$ cm$^{-2}$ that favors contact formation and electrical transport studies. The Wannier orbitals for the first and second dispersive moiré valence bands are localized at the MoTe$_2$ and WSe$_2$ layers, respectively. They form a triangular lattice structure in each layer, which, once brought near resonance by the displacement field, hybridize to realize an effective honeycomb lattice described by an extended Kane–Mele–Hubbard model [25] (Figure 1(a)). An electric field perpendicular to the sample plane can continuously tune the degree of band inversion (or sublattice potential) through the Stark effect. The moiré bands are non-topological and topological before and after band inversion, respectively. From now on, we will focus on half-band-filling of the first moiré valence band, corresponding to a filling factor of one hole per moiré unit cell ($\nu = 1$).

## II. TOPOLOGICAL PHASE TRANSITIONS

Figure 1(b) shows the schematic of our dual-gated device of MoTe$_2$/WSe$_2$ employed in this study (see Supplementary Materials for device fabrication details [55]). The top and bottom gate voltages ($V_t$ and $V_b$) independently control the hole filling factor ($\nu$) and the vertical electric field ($E$). $\nu = 1$ is defined as one hole per moiré unit cell, corresponding to a half-filled moiré valence band. All transport and optical data are obtained from the same devices as reported in our earlier study (device 1) [26], unless otherwise specified.

Figure 1(c) shows the longitudinal resistance ($R_{xx}$) versus ($\nu$, $E$) at zero magnetic field ($B = 0$ T). The QAH state is evidenced by the $R_{xx}$ dip [26], which is consistent with the MCD hotspot in Figure 1(d). Figure 1(f) summarizes the electric field ($E$) dependent $R_{xx}$ at $B = 0$ T and varying temperatures ($T$). The corresponding temperature dependent $R_{xx}$ at selected $E$-fields is shown in Figure 1(g). Three regions can be identified. $R_{xx}$ Increases quickly with decreasing temperature for $E \lesssim 0.68$ V/nm; this region corresponds to a Mott (or charge-transfer) insulating state before band inversion. Holes are located only in the MoTe$_2$ layer, forming a triangular lattice, i.e., a Mott insulator (Figure 1(a), left panel). After band inversion, this state transitions to a QAH insulator for $0.685 \lesssim E \lesssim 0.705$ V/nm; here. $R_{xx}$ drops rapidly below the magnetic transition temperature and a quantized Hall resistance, $R_{xy} = h/e^2$, emerges due to chiral edge state transport ($h$ and $e$ denote the Planck's constant and the electron charge, respectively). Holes occupy both TMD layers and form a honeycomb lattice [24]( Figure 1(a), middle panel). Upon further increase in $E$ beyond $\approx 0.705$ V/nm, a new correlated insulating state with diverging $R_{xx}$ at low temperatures emerges (which is suppressed under even higher $E$-field, see Fig. S1). We will show below that this state is an antiferromagnetic (AFM) Mott insulator (Figure 1(a), right panel).

At the first phase boundary separating the Mott and QAH insulator at $E \approx 0.685$ V/nm, the electric field dependence of $R_{xx}$ does not exhibit a single crossing point as temperature varies. This is consistent with the reported topological phase transition between the two states without charge gap closure [26,34]. The transition is likely a weak first-order phase transition with a small gap discontinuity, which could be smeared out by disorder broadening in our devices [26]. In contrast, a clear single crossing point is observed for the $R_{xx}$ isotherms at the second phase boundary. It corresponds to a nearly temperature independent $R_{xx} \approx 10$ kΩ at the critical electric field $\approx 0.705$ V/nm as demonstrated in Figure 1(g).

Here the system is metallic with $R_{xx}$ extrapolating to a finite value in the zero-temperature limit. This critical metallic state separates the QAH and VC-AFM insulators [3,4,35]. These observations are consistent with a continuous TPT accompanied by charge-gap closure. The continuous TPT is also supported by the absence of hysteresis in the MCD vs. $E$ curve (Fig. S6).

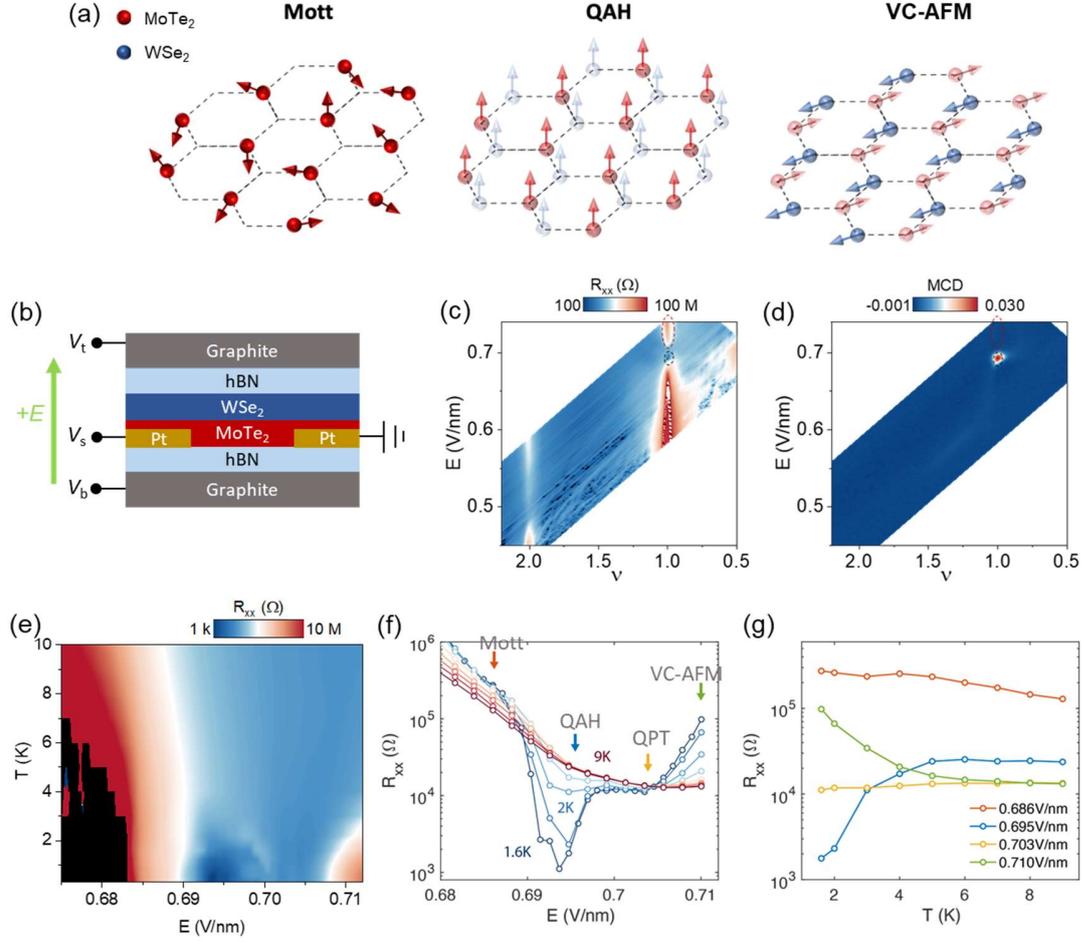

FIG. 1. Electric-field-tuned continuous quantum phase transition. (a) Schematics spin/valley/layer configurations of Mott insulating state, QAH, and VC-AFM states with increasing electric field at $\nu=1$. Red and blue balls represent holes in the $MoTe_2$ and $WSe_2$ layers, respectively. Since at high electric field, holes are shared between the two layers, the strong/light color of the balls corresponds to a large/small fraction of the holes residing in that layer. At the Mott insulator state, all charges are in the $MoTe_2$ layer, and spins are not ordered due to geometric frustration on a triangular lattice. With increasing electric field, a small fraction of holes is pushed from $MoTe_2$ to the $WSe_2$ layer and the Mott state transition to QAH, where both layers have spontaneous valley polarization and spins are co-aligned (out-of-plane spin). At even higher electric fields, more holes are pushed to the $WSe_2$ layer, and the QAH state undergoes a continuous quantum phase transition (QPT) into the VC-AFM state with collinear intervalley coherence but no spontaneous valley polarization (in-plane spin). (b) Schematic of the dual-gated device for optical and transport measurements. $V_t$, $V_b$ and $V_s$ are the bias voltages applied to the top/bottom graphite/hBN gate, and the channel source, respectively. (c),(d) Longitudinal resistance and MCD at 1.65 K as a function of ($\nu$, $E$). $R_{xx}$ is measured under zero magnetic field; MCD is the antisymmetrized result under an out-of-plane magnetic field of ±5 mT. The due-gated geometry of the devices enable varying the filling factor and $E$-field independently. The black and red dashed circles mark the region of the QAH insulator and the VC-AFM states, respectively. (e) $R_{xx}$ as a function of $E$ and $T$ at fixed doping $\nu = 1$. (f) Electric-field dependent $R_{xx}$ curves at varying temperatures. At $E \lesssim 0.685$ V/nm Mott insulating state, $R_{xx}$ increases quickly with decreasing temperature. It transitions to a QAH insulator for $0.685 \lesssim E \lesssim 0.705$ V/nm, where $R_{xx}$ drops rapidly below 6 K. At $E \gtrsim 0.705$ V/nm, a new correlated insulating state with diverging $R_{xx}$ at low temperatures emerges, which we identified as the VC-AFM state. A single crossing point of $R_{xx} \sim 10$ kΩ is observed at the critical metallic state on the phase boundary between QAH and VC-AFM, consistent with a continuous QPT. (g) Temperature dependence of $R_{xx}$ at selected electric fields. The selected electric fields are denoted by arrows in (f).

## III. MAGNETIC PROPERTIES

To better understand the nature of the different Mott states and the associated TPTs, we characterize the magnetic properties of the material using MCD spectroscopy, which measures the spin-valley polarization or magnetization of each individual TMD layer (see Supplementary Materials [55]). Figure 2(a) shows the magnetic field-dependent MCD spectrum at 1.65 K for the three different Mott states and the critical metallic state. The MoTe$_2$ layer is always heavily hole-doped so that the MCD spectrum is dominated by the attractive polaron resonance near 1.12 eV [36].

In contrast, the WSe$_2$ layer is charge-neutral for the Mott insulator state (before band inversion) and hole-doped for the rest (after band inversion); the neutral exciton and attractive polaron resonance dominate the MCD spectrum, respectively [37]. In Figure 2(b), we show the magnetic field-dependent MCD integrated over a narrow energy window near the WSe$_2$ resonances; it represents the magnetic field-dependent spin-valley polarization in the WSe$_2$ layer at 1.65 K. We further extract the MCD slope at zero magnetic field, which is proportional to the magnetic susceptibility ($\chi$) of the sample. The temperature-dependent susceptibility for all four states is shown in Figure 2(c); in the same plot, we also show the corresponding temperature-dependent $R_{xx}$ at $B = 0$ T.

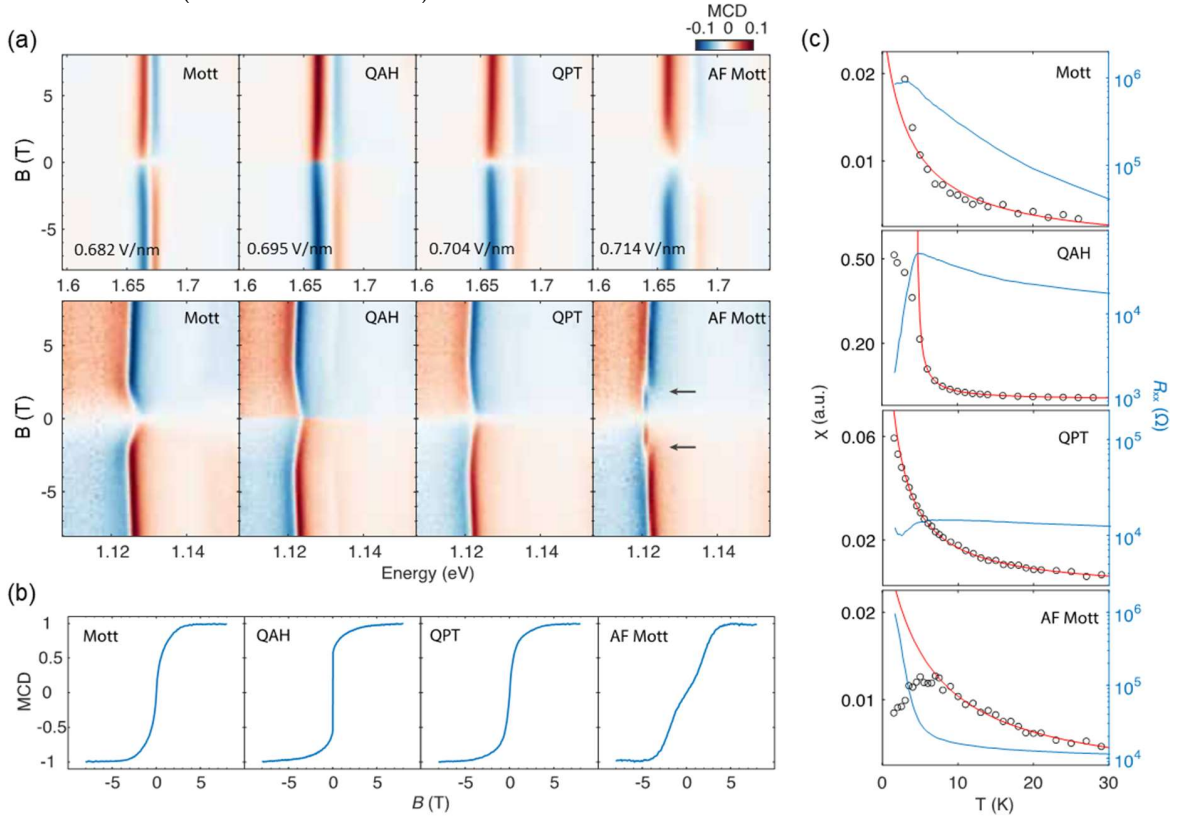

FIG. 2. Magnetic properties at varying electric fields. (a) Magnetic-field dependent MCD spectra at different electric fields for WSe$_2$ (top) and MoTe$_2$ (bottom). At the AFM Mott state, a spectral jump is observed near a critical magnetic field $B^*$, as marked by arrows. The critical magnetic field $B^*$ increases with increasing electric field. (b) Magnetic-field dependence of integrated MCD at different electric fields for WSe$_2$. The two resonances at low and high energy correspond to attractive and repulsive polarons in WSe$_2$, respectively. In the VC-AFM state, the attractive polaron shows a pronounced spectral shift with $B$ field below $B^*$. (c) Temperature dependence of magnetic susceptibility and $R_{xx}$ at varying electric fields. At the Mott insulator state, the susceptibility increases with decreasing temperature, showing paramagnetic behavior. At QAH, the susceptibility diverges at finite temperature, consistent with a ferromagnetic ground state. For VC-AFM, a broad susceptibility peak is observed at $T \approx |\theta|$; concurrently, $R_{xx}$ increases by orders of magnitude near $|\theta|$.

We first discuss the Mott insulator. The MCD resonances at 1.65 K evolve smoothly with magnetic field. The magnetic susceptibility follows a Curie-Weiss dependence, $\chi^{-1} \propto T - \theta$, at temperatures ($T$) higher than the Curie-Weiss temperature ($\theta$). $\theta \approx -5$ K is obtained at a representative $E$-field of 0.682 V/nm, providing a measure of the exchange energy scale between the local moments in the system, where the negative sign reflects AFM interaction. At 1.65 K, the integrated MCD scales linearly with the magnetic field, without the suppression of slope at small field; the behavior is consistent with that of a paramagnet. The paramagnetic behavior at temperatures substantially smaller than $|\theta|$ reflects the effects of geometric frustration in a triangular lattice, and the ground state is expected to be a 120-degree Néel AFM order [38,39]. The geometric frustration and the 2D nature of the material are expected to greatly suppress the ordering temperature from $|\theta|$.

In contrast to the Mott insulator, the MCD resonances and the integrated MCD for the QAH insulator exhibit a jump at zero magnetic field at 1.65 K; the jump corresponds to the emergence of spontaneous spin-valley polarization. The high-temperature magnetic susceptibility follows the Curie-Weiss law; the susceptibility diverges at $\theta \approx 4$ K, below which spontaneous spin-valley polarization develops; the positive $\theta$ reflects the ferromagnetic (FM) exchange interaction between the local moments; the observations are consistent with a FM ground state as reported by a recent study [26].

Ferromagnetism vanishes at the critical metallic state, which shows a smooth magnetic field dependence for the MCD at 1.65 K with magnetic saturation at ~ 1 T. A Curie-Weiss temperature $\theta \approx 0$ K can be extracted, consistent with a paramagnetic state. Lastly, we discuss the new insulating state that emerges at the highest electric fields. In contrast to all other cases, the MCD resonances exhibit a sudden spectral shift at a characteristic magnetic field $B^*$, which increases with the electric field (Fig. S2). The spectral shift is less visible in the spectrum of WSe$_2$, likely due to its much broader polaron resonance. Correspondingly, the integrated MCD increases with magnetic field with a suppressed slope for $B < B^*$; the slope increases near $B^*$ before magnetic saturation. High-temperature Curie–Weiss fitting at a representative $E$-field of 0.714 V/nm yields $\theta \approx -5$ K, evidencing predominant AFM exchange. In addition, there is a broad susceptibility peak at $T \approx 6$ K, indicating the AFM ordering temperature, i.e., Néel temperature $T_\text{N}$, in reasonable agreement with the extracted $|\theta|$. Below $T_\text{N}$, the zero-magnetic-field $R_{xx}$ increases quickly by orders of magnitude; this is in contrast to its weak temperature dependence above $T_\text{N}$. These observations suggest the emergence of an AFM-ordered Mott insulator below the transition temperature $T_\text{N}$. The decrease in $\chi$ below $T_\text{N}$ correlates with the emergence of a metamagnetic-like (field-induced reorientation) transition near $B^*$ at low temperatures [40]. The characteristic magnetic field $B^*$ is expected to be directly proportional to $|\theta|$ for an AFM system at temperatures small compared to $|\theta|$. This is supported by the similar electric field dependence for $|\theta|$ and $B^*$ (Fig. S2 and S4).

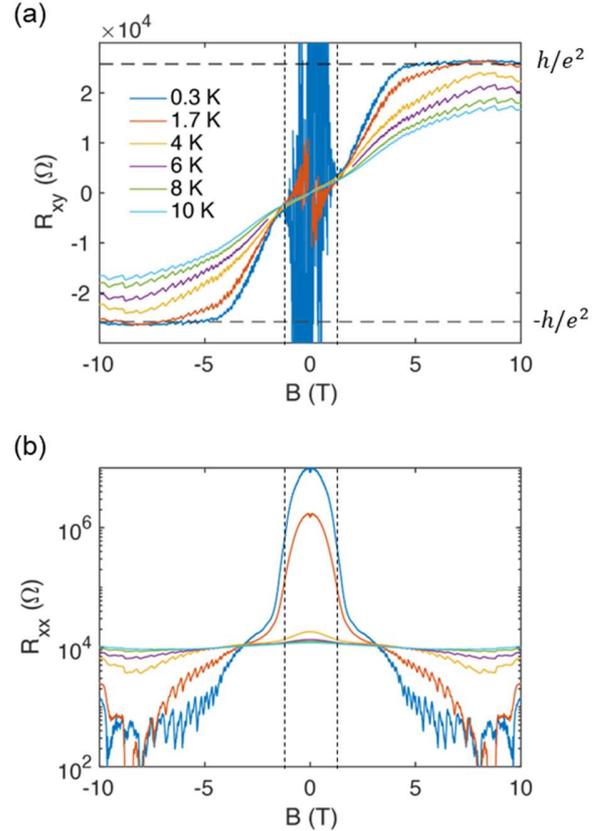

FIG. 3. Field-induced valley pseudospin polarization at VC-AFM. (a), (b), Magnetic field dependent $R_{xy}$ (a) and $R_{xx}$ (b) at varying temperatures in the VC-AFM state. At low temperature, $R_{xy}$ increases with increasing B field and saturates to $h/e^2$ at large magnetic field. A clear slope change is observed near critical field $B^* \approx 1.2\ T$, as indicated by the dashed lines. $R_{xx}$ decreases by orders of magnitude with increasing magnetic field, and nearly vanishes at large magnetic field. With increasing temperature, the saturation behavior of $R_{xy}$ disappears. The low temperature behavior is consistent with the valley pseudospin being polarized to the out-of-plane direction by increasing $B$ field, inducing a quantum

phase transition from the VC-AFM state to a Chern state.

The emergence of an AFM mott insulator at $T < |\theta|$ is further supported by the magneto-transport properties at the same electric field (Figure 3). Correlated with the MCD response, $R_{xy}$ increases with the magnetic field, and a clear slope change is observed near $B^*$; correspondingly $R_{xx}$ decreases by orders of magnitude, and nearly vanishes at large field (Because of the large $R_{xx}$ at low temperatures, the measurement of $R_{xy}$ is unreliable at small magnetic fields.) The results not only support the emergence of an AFM-ordered Mott insulator but also demonstrate a metamagnetic-like transition near $B^*$ from the AFM insulator to a Chern insulator with quantized Hall transport. With increasing temperature, the $R_{xx}$ kink and the MCD spectral shift near $B^*$ weaken (Fig. S3); the magnetic transition near $B^*$ disappears at $T \approx |\theta|$. The results are consistent with a thermodynamic phase transition to an AFM-ordered state at $T \approx |\theta|$. Also, similar to the electric-field-induced transition (Figure 1), there is a single crossing point of $R_{xx} \sim 10$ k$\Omega$ at $B^*$ for the isothermal magnetoresistance curves, i.e. $R_{xx}$ is nearly temperature independent at $B^*$. The result suggests a continuous field-induced magnetic transition [4,35]. Although electric-field-tuning of AFM-FM switching field has been demonstrated in 2D magnet CrI$_3$ [41,42],
and FM to paramagnetic Mott insulator with AFM exchange has been demonstrated in TMD moiré systems [15,26,43], a pure electric-field-induced AFM-FM transition has never been observed in moiré systems and any other material systems, to the best of our knowledge.

**IV. PHASE DIAGRAM AT HALF-FILLING**

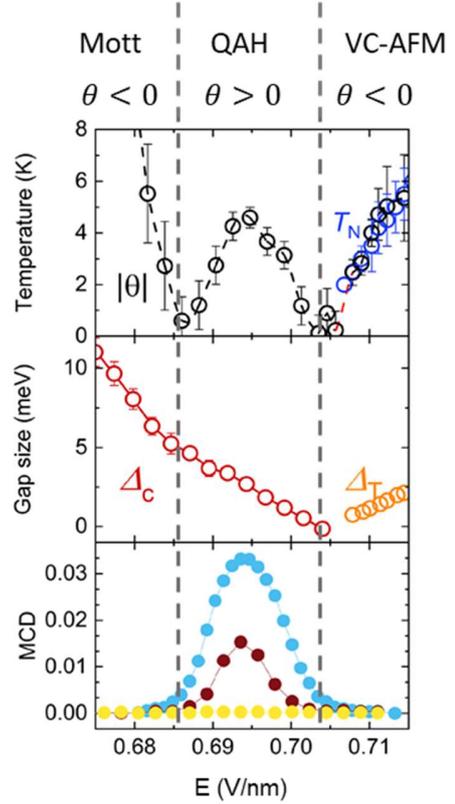

FIG. 4. Phase diagram at $\nu=1$. Top panel: electric field-dependent Curie-Weiss temperature $|\theta|$. The Curie-Weiss fitting of $\theta$ is shown in Fig. S4. Middle panel: Charge gap as a function of electric field extracted from capacitance measurements ($\Delta_c$) at 300 mK (red circles, device 2, the same device as reported in our earlier capacitance study [26]) and from thermal activation transport ($\Delta_T$, orange circles, device 1). See Fig. S5 for thermal activation fit to extract gap size. The charge gap continuously vanishes at $E_c \approx$ 0.704 V/nm that separates QAH and VC-AFM, and reopens above $E_c$. Bottom panel: spontaneous MCD as a function of electric field at 1.6 K (blue), 4 K (brown), and 7 K (yellow). A finite spontaneous MCD is observed in the QAH region.

We summarize our findings with a phase diagram in Figure 4. With increasing electric field, $\theta$ changes from negative (AFM) for the Mott insulator to positive (FM) for the QAH and back to negative for the VC-AFM insulator. Spontaneous valley polarization is observed only for the QAH insulator (i.e., the FM region). The charge gap obtained from compressibility measurements vanishes continuously at the critical electric field $E_c \approx 0.704$ V/nm that separates the QAH and AFM insulators. It reopens above $E_c$; the gap size for $E > E_c$ is determined by fitting the temperature-dependent $R_{xx}$ to a thermal activation behavior (Fig. S5). As reported by a recent study, the charge gap evolves smoothly across the boundary near 0.685 V/nm separating the Mott and QAH insulators [26]; the transition is likely a weak first-order TPT. The second TPT at $E_c$ is continuous.

## V. DISCUSSIONS

In our experiment, the electric field continuously tunes the interlayer/sublattice potential or the degree of band inversion. Holes form a triangular moiré lattice before band inversion. Band inversion occurs at the first TPT near 0.685 V/nm; after this point, holes are shared between the two TMD layers and form a honeycomb moiré lattice (Figure 1). With additional band inversion, the system continuously transitions from a QAH insulator to an VC-AFM insulator at $E_C$. The Mott insulator before band inversion is geometrically frustrated [38,39]; it most likely has non-collinear spin arrangements (Figure 1(a)). The QAH insulator is spin-polarized (probably with some degrees of spin canting) [37,44]. We propose that the AFM mott insulator for $E > E_c$ has a spin order shown in Figure 1(a) for the following reasons: The absence of a spontaneous spin-valley polarization in either of the TMD layers, as revealed by our layer-resolved MCD results (Figure 2), rules out a staggered Néel order of anti-aligned spins with non-zero out-of-plane component. The spins in each TMD layer can only be in-plane aligned; otherwise, a finite spontaneous MCD would have been observed. In other words, the AFM state can only be valley-coherent [45,46] but not valley-polarized in the constituent TMD layers. Moreover, because geometric frustration is absent in a honeycomb lattice, a collinear Néel order has a lower ground state energy than a noncollinear Néel order (e.g., 120-degree Néel order in a triangular lattice) [39,47]. Therefore, the most reasonable AFM state consistent with experiment is a staggered Néel order of in-plane spins schematically shown in Figure 1(a). Each layer is in an intervalley coherent FM state forming a triangular substructure [45,46] (i.e. spontaneous phase coherence between the K and K' valleys within each TMD layer); the nearest neighbor AFM coupling between the two triangular substructures produces an interlayer $\pi$ phase shift in the valley coherence, resulting in the proposed Néel order.

The proposed Néel order is consistent with several experimental observations. First, the transition to an intervalley coherent state in two dimensions is a broad Berezinskii-Kosterlitz-Thouless (BKT) transition; the observed broad susceptibility peak at $T \approx |\theta|$ is consistent with the BKT scenario [48,49]. However, the broadness of the transition can also be caused by disorder in the heterostructure from inhomogeneous strain and twist angle brought on by the stacking and fabrication process. Compared to the Mott insulator, the honeycomb lattice without geometric frustration helps stabilize the long-range Néel order below the BKT transition temperature. Second, a perpendicular magnetic field can polarize the in-plane spins to the out-of-plane direction when the Zeeman energy exceeds the AFM exchange interaction; this is consistent with the observed magnetic transition to a spin-polarized Chern state for $B > B^*$ (Figure 2 and 3). The Zeeman energy $g\mu_B B^* \approx 0.4 - 0.6$ meV is also comparable to the measured AF exchange energy scale $k_B|\theta| \approx 0.4$ meV [40]. Here we estimate $g_{eff} \approx 7 - 10$ [50–52] (see Supplementary Materials for details [55]), $g_{eff}$, $\mu_B$ and $k_B$ denote the effective hole g-factor, the Bohr magneton and the Boltzmann constant, respectively. Third, because the interlayer tunneling amplitude is spin-dependent, a spectral shift in the attractive polaron resonance is expected near the metamagnetic-like transition [36,53,54]; this is observed near $B^*$ in the MCD spectrum (Figure 2(a)).

## VI. CONCLUSIONS

We have mapped an electric-field–driven sequential topological phase transitions between three distinct correlated insulating phases at $\nu = 1$ in moiré MoTe$_2$/WSe$_2$. The second transition displays simultaneous charge-gap closure and magnetic reconstruction into a valley-coherent AFM Mott state, providing a rare platform to study correlation-topology intertwined quantum criticality. Future spectroscopic and scaling probes will be essential to establish the universality class of the critical metallic state and to search for fractionalized excitations.

ACKNOWLEDGMENTS

Research was primarily supported by the National Key R&D Program of China (Nos. 2021YFA1400100, 2021YFA1401400, 2022YFA1405400, 2022YFA1402702), the National Natural Science Foundation of China (Nos. 12550403, 12174250, 12141404, 12350403, 12174249, 92265102, 12374045), the Innovation Program for Quantum Science and Technology (Nos. 2021ZD0302600 and 2021ZD0302500), the Natural Science Foundation of Shanghai (No. 22ZR1430900). S.J. and T.L. acknowledge the Shanghai Jiao Tong University 2030 Initiative Program B (WH510207202). T.L. and S.J. acknowledge the Yangyang Development Fund. K.W. and T.T. acknowledge support from the JSPS KAKENHI (Grant Numbers 21H05233 and 23H02052) , the CREST (JPMJCR24A5), JST and World Premier International Research Center Initiative (WPI), MEXT, Japan.